\documentstyle[12pt]{article}

\begin{document}

\title
{
Monte Carlo Study of the Inflation-Deflation Transition
in a Fluid Membrane
}
\author
{
B. Dammann$^1$,
H.C. Fogedby$^2$,
J.H. Ipsen$^3$,
and C. Jeppesen$^4$\\
\\
$^{1,3}$Department of Physical Chemistry\\
The Technical University of Denmark\\
DK--2800 Lyngby, Denmark\\
\\
$^{2}$Institute of Physics and Astronomy \\
 University of Aarhus \\
 DK-8000 Aarhus C, Denmark \\
\\
$^{3}$Materials Research Laboratory\\
University of California, Santa Barbara \\
CA 93106, USA
}
\maketitle
\footnotetext[1]{bernd@membrane.fki.dth.dk, $^2$fogedby@dfi.aau.dk,\\
$^3$ipsen@lipid.fki.dth.dk, $^4$jeppesen@polymer.ucsb.edu}
Key words:
\noindent
Random surface, fluid membrane, dynamic triangulation, osmotic
pressure, branched polymer configuration, deflation, inflation,
first order phase transition, critical pressure,
finite size scaling, reweighting, histogram technique, scaling,
scaling exponents
\newpage

\def\baselinestretch{1.5}
\small\normalsize

\begin{abstract}
We study the conformation and scaling properties of a self-avoiding fluid
membrane, subject to an osmotic pressure $p$, by means of Monte Carlo
simulations. Using finite size scaling methods in combination with
a histogram reweighting techniques we find that the surface
undergoes an abrupt conformational transition at a critical pressure $p^\ast$,
 from low pressure deflated configurations
with a branched polymer characteristics to a high pressure inflated phase,
in agreement with previous findings \cite{gompper,baum}.
The transition pressure $p^{\ast}$ scales with the system size
as $p^\ast \propto N^{-\alpha}$, with $\alpha = 0.69 \pm 0.01$.
Below $p^\ast$ the enclosed volume scales as $V \propto N$,
in accordance with the self-avoiding branched polymer structure,
and for $p\searrow p^{\ast}$ our data are consistent with the finite size
scaling form $V \propto N^{\beta_{+}}$, where $\beta_{+} = 1.43 \pm 0.04$.
 Also the finite size scaling
behavior of the radii of gyration and the compressibility moduli
are obtained. Some of the observed exponents and the mechanism
behind the conformational collapse are interpreted in terms of
a Flory theory.
\end{abstract}

\newpage
\section{Introduction}

Over the past several years there has been an increased interest in the phase
behavior and morphological properties of flexible, fluid interfaces.
Beside the theoretical challenge in understanding surfaces as two
dimensional generalizations of polymers, they are expected to be of
relevance in physical systems ranging from simple interfaces between
coexisting fluids close to the consolute point
to surfactant interfaces in e.g. microemulsions and
lipid membranes\cite{meunier,weinberg,Dubois,Genn}.

In the context of modelling lipid vesicles it is of particular
importance to study the conformation of closed membranes in the
presence of surface tension, bending rigidity, osmotic pressure
between the interior and the exterior of the membrane, etc.
For a simple, closed, fluid membrane with
bending rigidity $\kappa\cdot k_B T$,  pressure $p\cdot k_B T$ and fixed
overall
topology, the vesicle conformation is determined by the
energy functional,
\begin{equation}
H/k_BT=\frac{\kappa}{2}\int dA(\frac{1}{R_1}+\frac{1}{R_2})^2-p\int dV,
\label{eq:H}
\end{equation}
where $R_1$ and $R_2$ are the principal radii of curvature \cite{helfrich}.
This phenomenological form is the basis for detailed descriptions
of shape transformations of rigid vesicles with $\kappa \gg 1$ \cite{deu,mia}.
There is a general consensus that sufficiently large, unconstrained
fluid self-avoiding surfaces collapse into branched polymer-like structures
with the characteristics of branched polymers for $T > 0 $\cite{Dur}.
This general feature will not be
changed by e.g., the presence of bending stiffness. However, the
osmotic pressure difference between the interior and exterior of a
closed surface is a quantity of direct experimental relevance,
which is expected to change this situation.

The properties of two-dimensional vesicles (ring-polymers), subject
to a pressure $p$ controlling the enclosed volume (area), have
been explored in a number of studies
\cite{leibler1}--\cite{gaspari}.
Leibler, Singh and Fisher \cite{leibler1}
verified by numerical studies that 2-D flexible vesicles change
continuously from a deflated state with characteristics of branched
polymers to an inflated state as $p$ is increased. This
change is accompanied by crossovers in scaling exponents characterizing
the surface. The scaling-behavior of the inflated regime can be
interpreted in terms of Pincus stretching exponents for polymers
\cite{maggs,pincus}.
Real-space-RG studies of a lattice model equivalent to the 2-D vesicle problem
\cite{ban91} and exact configurational enumeration techniques \cite{Fish}
show agreement with the numerical studies. The universal relations have
also
been confirmed by use of conformal techniques on the lattice model
\cite{Cardy}.
Furthermore, analytical studies of pressurized, flexible, but self-intersecting
surfaces in 2-D have been carried out \cite{gaspari}.
Only few studies have been carried out in 3-D.
In \cite{banavar,Stel1}  pressurized,
flexible, fluid surfaces have been considered in the context of
the confined phases of some lattice gauge models.
These studies have focussed on the effect of pressure on vesicle topology for
deflated vesicles.

It has recently been demonstrated by numerical studies
that pressurized, flexible surfaces
in 3-D undergo a first-order-like transition from a deflated phase
with branched polymer behavior to inflated configurations. Gompper and
Kroll based their studies on a triangulated random surface model
\cite{gompper},
while Baumg\"{a}rtner applied a plaquette model for a self-avoiding surface
\cite{baum,baum2},
where the surface is composed of the domain walls of a spin-model.
In these studies the basic scaling relations, established for 2-D flexible,
pressurized surfaces, were generalized to three dimensions.

In the present paper we also consider a fluid membrane with an osmotic
pressure $p$, but with zero bending rigidity $\kappa$.  Its properties
are analyzed on the basis of computer simulations of a
triangulated random surface representation of the fluid membrane.
Using the Ferrenberg-Swendsen
reweighting techniques \cite{ferrenberg} we attempt
a precise determination of the critical pressure and the scaling
exponents in the deflated and inflated phases.

\section{Model}
The model considered here is a simple extension of the
Ho-Baumg\"{a}rtner description of self-avoiding surfaces \cite{ho}.
It consists of a closed triangular network of beads
positioned at the $N$ vertices of the network with the topology of
a sphere. To each vertex position
$\vec{X}_i$ is associated a bead with hard core diameter $1$. In order
to enforce the self-avoidance constraint the length of the flexible
tether between neighboring vertices is chosen to be less than $\sqrt{2}$.
The fluidity of the surface is modelled
by a dynamical change of the connectivity between the vertices by
means of `flip' transformations which correspond to deleting
a tether or link between neighboring vertices and attempting to
form a new link between the adjacent vertices in the two
triangles involved.
Finally, the shape
changes are generated by a local updating of the vertex position, the
`shift' transformation, $\vec{X}_i\rightarrow\vec{X}_i+\delta\vec{X}_i$,
where $\delta\vec{X}_i$ represent an incremental change in
the local surface position. $\delta\vec{X}_i$ is picked randomly within
a cube, where the cube-side is adjusted to preserve self-avoidance.
 These procedures ensure sampling over all
possible self-avoiding, piece-wise linear surfaces consisting of $N$ beads
and with the topology of a sphere.  The updating of the surface shapes
is posed by standard Monte Carlo techniques \cite{ole} on the partition
function
\begin{equation}
Z = \sum_{Surface\  shapes} \exp(-H/k_BT)\,.
\label{eq:Z}
\end{equation}

In the simulations a discretized version of the Hamiltonian from
Eq.(\ref{eq:H}) with $\kappa = 0$ is applied.
In order to obtain a fast updating rate a dynamic sublattice structure
is implemented which keeps track of the $N$ vertices
during shape
transformations. We furthermore use data structures consisting
of linked lists which, given an arbitrary
vertex or link, allow  an identification of the adjacent
triangles.
The last feature implies an oriented triangulation  and leads to the following
expressions for the area of the surface and the enclosed volume:
\begin{equation}
A=\sum_{\rm n}\Delta A_{\rm
n}=\frac{1}{2}\sum_{ijl}\mid(\vec{X}_i-\vec{X}_j)\times
(\vec{X}_l-\vec{X}_j)\mid,
\label{eq:A}
\end{equation}
and
\begin{equation}
V=\sum_{\rm n}\Delta V_{\rm n}=\frac{1}{6}\sum_{ijl}\vec{X}_i\cdot
(\vec{X}_j\times
\vec{X}_l).
\label{eq:V}
\end{equation}
The indices $ijl$ pertain to the corners of an oriented triangle and the
sum runs over all triangles of the triangulation.

Using a stress ensemble we have collected data for a range of $p$
values and $N$ values ($N$ is the number of vertices). In a simulation
we typically measure the following quantities:
\begin{itemize}
\item We monitor the shift and flip rates in order to ensure a suitable
balance between fluidity (the flip rate) and shape changes
(the shift rate)

\item For the characterization of the over-all membrane size
 we  sample the radius of gyration $R_G$:
\begin{equation}
R^2_G=\frac{1}{N}\sum_{i=1}^{N}(\vec{X}_i-\vec{X}_{CM})^2,
\label{eq:I}
\end{equation}
where
$\vec{X}_{CM}=\frac{1}{N}\sum_{i=1}^N\vec{X}_i$ is the center of mass.

\item the area $A$ and the volume $V$ of the membrane are monitored.

\end{itemize}

The measured quantities are stored for construction of probability
distributions and evaluation of equilibrium thermal averages.
It is characteristic of fluctuating membrane systems that the correlations
times are quite long and grow fast with the system size. Careful
analysis of the time series is thus very important to ensure
proper equilibrium sampling of the fluctuating surface.
 E.g., for $N=400$ we sample 20 Mill MCS/vertex in order
to equilibrate the system and perform measurements over about
400 Mill MCS/vertex. We have investigated system sizes ranging from $N=49$
to $N=400$.

\section{Simulations and Data Analysis}
The Ferrenberg-Swendsen reweighting of the probability distributions
allow for an approximate determination of thermal averages for a
pressure value $p$ close to the simulation pressure $p_0$, if the phase space
is properly sampled, e.g. for the averaged volume $\langle V \rangle$ it takes
the form:
\begin{equation}
\langle V \rangle_p = \frac{\sum_i V_i P(V_i) \exp(-\Delta p V_i)}
{\sum_i P(V_i) \exp(-\Delta p V_i)},
\label{eq:<V>}
\end{equation}
where $\Delta p = p-p_0$, and $P(V_i)$ is the sampled distribution. The sum is
running
over all measured volume values.
Figure 1 shows the averaged vesicle volume versus pressure $p$ for
two different system system sizes $N=100$ and 196, obtained from simulations
at different pressures. The lines are the results of the reweighting technique,
using only one simulation near the assumed critical pressure.
The dramatic change in the vesicle volume over a narrow pressure
interval indicates the presence of a phase transition in the thermodynamic
limit $N \rightarrow \infty $
from a low pressure deflated phase to a high pressure inflated phase
in accordance with the simulations of Gompper and Kroll \cite{gompper}.
The plots of volume versus pressure thus allow an approximate
determination of the critical pressure $p^\ast$.
The reweighting technique also enables us to reconstruct the
phase diagram. In the insert of Fig. 1 we show the reconstructed
volume versus pressure
curves for $N=81,100,121,144,169,196,225,256,324,400$ obtained by use of
Eq.(\ref{eq:<V>}).

A more precise determination of the critical pressure can be obtained
by evaluating the total volume fluctuations $\langle V^2 \rangle
- \langle V \rangle^2$ and estimate the peak position. Figure
2 shows the measured $p^\ast$ for different system sizes. $p^\ast$
is well described by a power-law in $N$:
\begin{equation}
p^\ast \sim N^{-\alpha},
\label{eq:past}
\end{equation}
where $\alpha = 0.69 \pm 0.01$, obtained from a linear fit to the
double-logarithmic representation of the
data. All indicated error bars are related to the regression analysis.

 The Ferrenberg-Swendsen reweighting technique \cite{ferrenberg}
can also be applied for the
construction of the probability distribution $P(V)$ at specific
pressures near the anticipated critical pressure $p^{\ast}$. In Fig. 3
we show examples of measured histograms representing $P(V)$ and $P(R^2_G)$
for $N=144$ at $p=0.52$ and $p=0.54$. The measured
$P(V)$ at $p=0.52$ is in good agreement with the probability distribution
obtained from reweighting the histogram measured at $p=0.54$.
At $p^{\ast}$ $P(V)$ has a characteristic double-peak structure
representing the two phases. Lee and Kosterlitz used this technique
to determine finite-size transition points for lattice models by matching
the peak-heights, and by investigation of their finite-size behavior
information about the nature of the transition in the thermodynamic limit
can be obtained \cite{lee}. It is a key assumption in this procedure that the
probability distribution for some density can be well described as the
superposition of Gaussian distributions. This is not fulfilled for $P(V)$,
but some properties of the distribution will be considered. $P(V)$ can in a
first approximation be written as the superposition of contributions from
inflated and deflated contributions:
\begin{equation}
 P(V) \simeq a_{-} P_{-}(V) + a_{+} P_{+}(V),
\label{eq:P(V)}
\end{equation}

\noindent
as suggested by e.g., Fig. 3.
 The maximum probability volumes for the two peaks $V_{-}$ and $V_{+}$
can easily be estimated and is shown in Fig. 4. We find the dependence:
\begin{equation}
        V_{\pm} \sim N^{\beta_{\pm}},
\label{eq:Vpm}
\end{equation}
where the associated scaling exponents are
$\beta_{-}= 0.99 \pm 0.01$
and $\beta_{+}= 1.43 \pm 0.04$.

Since the two peaks in $P(V)$ display very different dependences of $N$,
 $P_{-}$ and $P_{+}$ have been analyzed separately.
In Fig. 5 $P_{+}$ is given in a double logarithmic representation up to a
multiplicative factor.
The plot indicates that the volume distribution for the inflated phase
of the surface has a universal form for large $N$.
\begin{equation}
 P_{+}(V)\simeq \frac{1}{N^{\beta_{+}}}f_{+}(\frac{V}{N^{\beta_{+}}}),
\label{eq:P-+}
\end{equation}

\noindent
where $f_{+}$ is represented in Fig. 5 as the limiting distribution as $N$
becomes large.  Our data are not sufficient for a thorough analysis
of the form of $f_+$.

 The probability distribution describing the volume fluctuations
of the deflated configurations are given in the insert of Fig. 5.
 Similar to $P_+$,
$P_-$  approach a universal form as $N$ becomes large:
\begin{equation}
P_-(V) \simeq  f_-(V-V_{-}),
\label{eq:P--}
\end{equation}

\noindent
$f_-$ is represented in the insert. The volume fluctuations
$\langle (\Delta V)^2 \rangle_{-}$ are thus independent of $N$.

At $p^\ast$ the probability distribution for the radius of gyration
$R^2_G$ displays a double-peak similar to the probability distribution
of the volume, as shown in Fig. 3.
The peak-positions are identified with the radii
of gyration of the inflated and the deflated phases $R^2_{\pm}$.
The calculated  $R^2_{\pm}$  shown in Fig. 6 indicate
a relationship:
\begin{equation}
    R^2_{\pm} \sim N^{2 \nu_{\pm}},
\label{eq:R2}
\end{equation}
where $2 \nu_- = 1.05 \pm 0.05$ and $2 \nu_+ =0.81 \pm 0.01$.

In Fig. 7 the data for the total volume fluctuations
$K =  \langle V^2 \rangle - \langle V \rangle^2$  at $p^\ast$ is plotted
against $N$. From the regression analysis in the log-log representation
we obtain a dependence:
\begin{equation}
 K(p^\ast ) \sim N^\gamma,,
\label{eq:Ctot}
\end{equation}

\noindent
where $\gamma = 3.62 \pm 0.02 $.

\section{Discussion and Conclusion}
 The above analysis confirms that fluid, flexible vesicles undergo
an abrupt conformational change between a deflated and an inflated
regime at some small pressure value which diminishes with the system
size according to a power-law, $p^{\ast} \propto N^{-\alpha}$, where
$\alpha = 0.69 \pm 0.01$. The value of this exponent is
in disagreement with previous findings 0.5 \cite{gompper} and
1 \cite{baum}.

In the deflated regime $p<p^\ast$ the data are consistent with branched
polymer configurations of the surface:
\begin{equation}
 \langle V \rangle_{-} \propto \langle R_G^2 \rangle_{-} \propto N,
\label{eq:VRN}
\end{equation}
and thus support the general expectation that unpressurized, fluid
membrane conformations are
controlled by the branched polymer fixed point, i.e., the stable
$\kappa =0$ crumpling fixed point \cite{gompper,boal}.

In the inflated phase for $p>p^\ast$, generalizing Pincus' result for
polymer stretching \cite{pincus}, Gompper and Kroll \cite{gompper}
and Baumg\"{a}rtner \cite{baum} propose the scaling form
\begin{eqnarray}
\langle R^2_G \rangle_{+}
&\sim& p^{2(2\bar{\nu} -1)}N^{2\bar{\nu}} \nonumber\\
\langle V \rangle_{+} &\sim& p^{3(2\bar{\nu} -1)}N^{3\bar{\nu}},
\label{eq:VR}
\end{eqnarray}
\noindent
where $\bar{\nu}$ is a new scaling exponent characterizing the stretched phase.
In the following we will use the short notation
$R=\sqrt{\langle R^2_G \rangle_{+}}$.

At the critical pressure inserting the scaling result $p^\ast\sim N^{-\alpha}$
in Eq. (\ref{eq:VR}), we obtain for $\langle V \rangle_{+}$ in the
inflated phase,
\begin{equation}
\langle V \rangle_{+} \sim N^{3[\bar{\nu}-\alpha (2\bar{\nu} -1)]},
\label{eq:V_}
\end{equation}
\noindent
and we derive the relationship
$\beta_{+}=3[\bar{\nu} -\alpha (2\bar{\nu} -1)]$. From the numerically
obtained exponents
$\beta_{+} \simeq 1.43$ and $\alpha\simeq 0.69$ in Eqs.(\ref{eq:Vpm})
and (\ref{eq:past}) we find
$\bar{\nu}\simeq 0.56$
which is in a reasonable agreement with Gompper and Kroll \cite{gompper}
($\bar{\nu} = 7/12=0.583$). However, a similar analysis of $R^2_G$
leads to a significantly different estimate $\bar{\nu} \simeq 0.75$.

Another consequence of Eq.(\ref{eq:VR}) is, that for
$p \searrow p^{\ast}$:
\begin{equation}
\frac{ \langle (\Delta V)^2\rangle_+}{\langle V \rangle_+}
=\frac{1}{\langle V \rangle_+} \frac{\partial \langle V \rangle_+}{\partial p}
 \sim (p^{\ast})^{-1} \sim N^{\alpha} \,,
\label{eq:Vrat}
\end{equation}
\noindent
which is contradictory to the numerical finding
$\langle (\Delta V)^2\rangle_{+}/\langle V \rangle_{+}
\sim N^{\beta_{+}}$, where $\langle (\Delta V)^2\rangle_{+} \sim
N^{2\beta_{+}}$
at the transition is a consequence of Eq.(\ref{eq:P-+}).
 Finally, we find
$\langle V \rangle_+ /R^3 \propto N^{\beta_+ - 3 \nu_+}$,
where $\beta_+ - 3 \nu_+ \simeq 0.22$ at
the transition, so the relation
$\langle V \rangle_+ \propto R^3$ is not consistent with our data.
The numerical data, obtained in the transition region, are not sufficient for
an extended scaling analysis, however, with the above considerations we
conclude that they do not support an ansatz of the type Eq.(\ref{eq:VR}).\\

 In the following we will discuss the inflation-deflation
collapse transition of a fluid surface within the framework of a Flory
theory which has been applied to the description of polymeric,
flexible surfaces \cite{Kantor} and attempted applied in the context
of plaquette surfaces \cite{Maritan84}.  The application of sufficient
pressure will restrict the configurational phase space
and prevent branching of the surface. Such a surface can for "small"
pressures allow an effective description in terms of a Flory theory
for the free energy:
\begin{equation}
F_{i} = \frac{R^2}{R_o^2} + \upsilon \frac{N^2}{R^d}
        -p \gamma N R.
\label{eq:Fi}
\end{equation}

\noindent
The first two terms represent the standard Flory theory
for a surface, where $R_o$ is the radius of gyration for
a self-intersecting surface, $R_o \sim N^{\nu_o}$, where
$\nu_o = 0$, and $\upsilon$ is the excluded
volume parameter. $d=3$ is the dimension of the embedding space.
In the last term the pressure couples
to the volume:
\begin{equation}
 \langle V \rangle_+ \propto
\langle \int_{A} dA\ \vec{n}\cdot(\vec{X}-\vec{X}_{\rm CM})\rangle_+
        \propto \langle \cos{\theta}\rangle_+ \cdot N R,
\label{eq:Vfl}
\end{equation}

\noindent
where $\theta = \angle(\vec{X}-\vec{X}_{\rm CM}, \vec{n})$ and $\vec{n}$ is the
the surface normal at surface position $\vec{X}$. Here it
is used that $|\vec{X}-\vec{X}_{\rm CM}|$ is distributed around $R$ in
the inflated configurations, so Eq.(\ref{eq:Vfl}) represents the leading
contribution
to $\langle V \rangle_+$.
 The parameter $\gamma $ in Eq.(\ref{eq:Fi}) is thus proportional to
 $\langle \cos{\theta} \rangle_+  > 0$.
The stationary condition for $F_{i}$ at $p=0$ can be characterized by
a critical exponent $\nu$:
\begin{eqnarray}
R &\propto& N^{\nu}, \ \ \ \nu = \frac{2}{d+2}=\frac{2}{5}=0.4, \nonumber \\
\langle V \rangle_+ &\propto& N R \propto N^{\beta},
\ \ \ \beta=1+\nu=\frac{7}{5}=1.4.
\label{eq:nu}
\end{eqnarray}

\noindent
These Flory exponents are indeed
close to the numerically obtained $\nu_+ = 0.41$ and
$\beta_+ = 1.43$. But since $F_{i} \propto N^{2 \nu}$,
these configurations are unstable with respect to
branched polymer collapse for large $N$ and $F_{i}$ must be
considered as a metastable free-energy branch for $p=0$.
 The branched polymer configurations are completely entropy
dominated with an approximate free energy \cite{gompper,Dur}:
\begin{equation}
 F_{bp} = -\ln (z^*) N + \frac{3}{2} \ln(N) - p N,
\label{eq:Fbp}
\end{equation}

\noindent
where $z^* > 1$.
Let us now apply a small pressure, so that the relations in Eqs.(\ref{eq:nu})
still hold. The characteristic pressure $p^{\ast}$ separating the
unstable $p=0$ region from the inflated region in $F_{i}$ is
given by $p^{\ast} N^{1+\nu} \sim N^{2 \nu}$ leading to
$ p^{\ast} \propto N^{-\bar{\alpha}}$, where
$\bar{\alpha}=1-\nu=3/5=0.6$. For a choice of $\bar{\alpha}$ smaller
than $3/5$ the perturbative considerations break down and the
pressure term will dominate in Eq.(\ref{eq:Fi}), which eventually
 leads to complete
inflation with $F \propto -pN^{3/2}$ for $N \rightarrow \infty$.
For $\bar{\alpha}$ larger than $3/5$ the surface will be controlled by
the $p=0$ behavior as $N$ becomes large.
This simple analysis thus suggests that the
trigger of the inflation-deflation
transition can be understood on basis of the cross-over from
the flaccid to the inflated conformations of closed, flexible,
polymeric surfaces under pressure. The strong
fluctuations in the transition region,
e.g., observed in the probability distributions, are
not included in this consideration. In particular
it must be expected that fluctuations in the weakly inflated
regime will modify the conditions for the stability of the inflated
configurations.
The deviation of $\bar{\alpha}$ from the numerically
obtained exponent $\alpha = 0.69$ is thus not surprising
in light of the crudeness of the approximations involved.

In the present paper we have reanalyzed the deflation-inflation transition
of a fluid membrane subject to an inflating pressure discovered by
Gompper and Kroll \cite{gompper}, and Baumg\"{a}rtner \cite{baum}.
 In the transition region we find low-pressure configurations
with the characteristics of branched polymers and an exponent for
the radius of gyration $\nu_-=\nu_{BP}=0.52 \pm 0.03$ and high pressure
configurations, with a characteristic exponent $\nu_+ = 0.41 \pm 0.01$.
Analysis of the volume distribution functions show that the two
phases display universal properties at $p^\ast$, characterized
by exponents for the volume $\beta_- = 0.99 \pm 0.01$ and
$\beta_+ = 1.43 \pm 0.04$ for the two types of configurations in the
transition region. Further, a simple
analysis of the transition within the framework of Flory theory
accounts for some of the observed exponents and suggests a
mechanism for the transition.
The exponents
characterizing the finite-size behavior of the transition pressure
and the total volume fluctuations are subjects to further investigations.\\
\\
{
\large
\noindent
Acknowledgements
}
\\
We would like to thank Ole G. Mouritsen and Martin J. Zuckermann
for illuminating discussions. We also wish to thank Benny Lautrup
for permission to run some of the simulations on his workstation.

\newpage

\noindent
{\bf Figures}

\begin{enumerate}
\item In this figure the volume averages $\langle V \rangle$
obtained from simulations at different pressures are plotted
vs $p$, for two different system sizes ($N=100$ and $N=196$).
The lines are the results from the Ferrenberg-Swendsen
reweighting for a single simulation at $p=0.7$ for $N=100$
and $p=0.445$ for $N=196$. The insert shows the volume
values obtained with the Ferrenberg--Swendsen reweighting technique
from simulations near the critical pressure, for $N=$ 81, 100, 121,
144, 169, 196, 225, 256, 324, 400.

\item This figure shows $\log(p^{\ast})$ vs $\log(N)$. The
values for $p^{\ast}$ are obtained from the peak in the total
compressibility $\langle (\Delta V)^2 \rangle$, which was calculated
using the Ferrenberg-Swendsen reweighting technique. The exponent
is obtained from fitting to $-0.69 \pm 0.01$.

\item The distributions $P(V)$ and $P(R^2_G)$ are plotted for
two different pressure values and the system size $N=144$. The
sharp peak in the $P(R^2_G)$ distribution belongs to the
inflated phase, whereas the corresponding peak in the $P(V)$
distribution is broader. For the branched polymer configurations it
is opposite.

\item The volumes $V_{-}$ and $V_{+}$ corresponding to the two
peaks in $P(V)$ at $p^{*}$ are plotted against $N$ in a
double logaritmic representation. The regression lines correspond
to $V_{\pm} \propto N^{\beta_{\pm}} $, where $\beta_{-} = 0.99 \pm 0.01$
and $\beta_{+} = 1.43 \pm 0.04$.

\item The unnormalized volume probability distrubutions at $p^\ast$
 for $N=$ 100, 121, 144, 169, 196, 225, 256, 324, 400. $P(V)$
 is for each $N$ multiplied by a common factor so the maxima are equal
 unity.
$P_+$: The part of the histograms associated with the inflated configurations
is plotted against $V/N^{\beta_+}$
in a log-log representation.
$P_-$: In the insert is the low-volume parts of the histograms
representing the inflated configurations similarly given in a
semi-logarithmic plot.

\item The radii of gyration $R_{-}^2$ and $R_{+}^2$ at $p^\ast$ are
plottet against $N$. The data are indicative of a relationship
$R^2_{\pm} \propto N^{\nu_{\pm}}$ where $\nu_{-}=1.05 \pm 0.05$
and $\nu_{+}=0.81 \pm 0.01$

\item The maximum of the total volume fluctuations
 $\langle (\Delta V)^2 \rangle = \langle V^2 \rangle  - \langle V \rangle^2$
 (corresponding to $p^\ast$) is plotted vs. $N$ in a log-log representation.
 The fit indicates $\langle (\Delta V)^2 \rangle_{\rm max} \sim N^{\gamma}$,
 with $\gamma = 3.62 \pm 0.02$
\end{enumerate}
\end{document}